\documentstyle[preprint,prd,eqsecnum,aps]{revtex}
\tighten

\newcommand{\beq}{\begin{equation}}
\newcommand{\beqa}{\begin{eqnarray}}
\newcommand{\eeq}{\end{equation}}
\newcommand{\eeqa}{\end{eqnarray}}
\newcommand{\simg}{\gtrsim}
\newcommand{\siml}{\lesssim}

\begin{document}

\draft
\preprint{UTAP-307}

\title{
Numerical Study of Inhomogeneous Pre-Big-Bang \\
Inflationary Cosmology
}

\author{Takeshi Chiba\footnote{Electronic address:
chiba@utaphp1.phys.s.u-tokyo.ac.jp, 
JSPS research fellow.}
}

\address{
Department of Physics, University of Tokyo, 
Tokyo 113-0033, Japan
}

\date{\today}

\maketitle

\begin{abstract}
We study numerically the inhomogeneous pre-big-bang inflation in a 
spherically symmetric space-time. We find that a large initial
inhomogeneity suppresses the onset of the pre-big-bang inflation. 
We also find that even if the pre-big-bang inflationary stage is
realized, the initial inhomogeneities are not homogenized.  
Namely, 
during the pre-big-bang inflation ``hairs''(irregularities) do not fall, 
in sharp contrast to the usual (potential energy dominated) inflation 
where initial inhomogeneity and anisotropy are  damped and
thus the resulting universe is less sensitive to initial
conditions. 

\end{abstract}


\section{introduction}

The pre-big-bang(PBB) cosmological scenario inspired by string theory 
has been suggested as a possible
working mechanism of inflation within the framework of the low energy
effective action of string theory\cite{pbb}.
In this scenario, the inflation occurs by the kinetic energy of a
massless dilaton, and the curvature and the string coupling grow until 
the effective action breaks down. After that epoch stringy
nonperturbative effects become important and hopefully enable the
universe to make a smooth transition to a standard
Friedmann-Robertson-Walker(FRW) epoch of decelerating expansion (for
recent attempts toward graceful exit, see \cite{exit}).

One of the aims for the introduction of the inflationary universe
scenario\cite{inf} was to solve the homogeneity problem. 
However, most of works on the PBB inflation is done within
the context of a homogeneous space-time (however, see \cite{inhomo}).
So, questions arise as to a homogenization process due to the PBB
inflation: 
Does the PBB inflation really homogenize initial inhomogeneities?
Can initial inhomogeneities prevent the onset of the PBB inflation?
The purpose of this paper is to answer these questions in a spherically
symmetric space-time. 

The peculiarity of the PBB inflation is that even classically the 
weak energy condition\cite{he} is violated during that epoch when the
equation are written in the form of the Einstein equations\cite{kar},
while the strong energy condition is violated but 
the weak energy condition is respected classically in the case of the
usual (potential energy dominated) inflation. This suggests that the
behaviour of inhomogeneities in the PBB inflation is quite different
from that in the usual inflation.
 
According to the linear perturbation analysis, 
the density perturbation $\delta$ of $p=w\rho$ matter, with $w$ being
a constant, behaves as\cite{ks} (in the long wavelength limit) 
$\delta \propto t^{{2(1+3w)\over 3(1+w)}}$. 
Similar behaviour is found by the approximation method to describe the 
super-horizon scale inhomogeneity (the gradient expansion
method)\cite{td}. Therefore, for the strong energy violating (but the
weak energy condition respecting) matter ($-1<w< -1/3$), the density 
perturbation decays with time. An inflaton generically satisfies this 
property. This behavior is related to the so-called cosmic no
hair conjecture, which states that all expanding universe  models 
with positive cosmological constant asymptotically approach the de
Sitter space-time\cite{gibbons}. 
On the other hand, however, 
for the weak energy condition violating matter ($w< -1$), 
the density perturbation {\it grows} with time.

Of course, the dilaton  does not have a constant $w$ and so the
above consideration is not directly related to the PBB inflationary 
cosmology, but it strongly suggests that the behavior of the density
perturbation in the PBB inflation is quite different from that in the
usual inflation. In order to study how  initial inhomogeneity affects
the onset and the duration of the PBB inflation, we need to consider
large inhomogeneities, and hence we have to resort to fully numerical 
treatment. In this paper we investigate the inhomogeneous
pre-big-bang inflation in a spherically symmetric space-time by solving 
the field equations fully numerically. A related numerical work based on 
a spectral method was appeared \cite{inhomo2}, 
however, the analysis is limited to weakly nonlinear 
perturbations. There also appeared a related analytical work
\cite{inhomo3} where a criterion for PBB inflation is proposed, which is
consistent with our numerical results. 

This paper is organized as follows. 
In Sec.2, basic equations based on the ADM formalism are given. 
In Sec.3, after the details of the initial condition and the numerical
method is described, 
the numerical results are given. Sec.4 is devoted to summary.

\section{field equations}

The low-energy effective action is given by
\beq
S=\int d^4x \sqrt{-g}{e^{-\phi}\over l_s^2}\left(~^{(4)}R +
g^{\mu\nu}\nabla_{\mu}\phi\nabla_{\nu}\phi\right),
\label{action}
\eeq
where $l_s$ is the string scale which is of order the
Planck scale, $\phi$ is the dilaton, and $~^{(4)}R$ is the four 
dimensional Ricci scalar. 
We have omitted other matter degrees of freedom for simplicity. 
Varying the action by ${g}^{\mu\nu}$ and $\phi$ yield
\beqa
G_{\mu\nu}&=&-\nabla_{\mu}\nabla_{\nu}\phi+{1\over 2}g_{\mu\nu}(\nabla
\phi)^2\equiv T_{\mu\nu},\label{einstein}
\\
\Box \phi&=& (\nabla \phi)^2,\label{scalar}
\eeqa
where $\nabla_{\mu}$ and ${\Box}$ are a covariant derivative and the 
d'Alembertian of ${g}_{\mu\nu}$, respectively. 

Although after the conformal transformation such that
$g_{E\mu\nu}=e^{-\phi}g_{\mu\nu}$ the action is reduced to that of 
a massless scalar field coupled minimally to the Einstein gravity, 
we shall work 
consistently in the string frame. This is because the gauge condition 
in the string frame metric has a direct geometrical meaning and
because the interpretation of the numerical results is direct and
clear as suggested in \cite{inhomo2}.

\subsection{Basic Equations in (3+1) Form}

In the ADM formalism, the line element generally takes of the form
\beq
ds^2=-\alpha^2dt^2 +\gamma_{ij}(\beta^idt+dx^i)(\beta^jdt+dx^j),
\eeq
where $\alpha,\beta^i$ is the lapse function, the shift vector,
respectively. $\gamma_{ij}$ is the three-metric of the spacelike 
hypersurface. A timelike vector
$n_{\mu}$ normal to the hypersurface is given as 
\beq
n_{\mu}=(-\alpha,0,0,0).
\eeq 
The extrinsic curvature $K_{ij}$  is then defined by 
\beq
K_{ij}=-{1\over 2}{\pounds}_n\gamma_{ij},
\eeq
where ${\pounds}_n$ denotes the Lie derivative along the integral curve
of $n^{\mu}$. 
The field equations Eq.(\ref{einstein}) then break up into evolution
equations 
\beqa
 \gamma_{ij,t}-{\pounds}_{\beta}\gamma_{ij}
& =& -2 \alpha K_{ij},\\
 K_{ij,t}-{\pounds}_{\beta}K_{ij}
& = &-{D_i}D_j\alpha + \alpha \left[R_{ij} +K K_{ij}-2K_{il}{K^l}_{j} 
 - \left(S_{ij} 
  + \frac{1}{2} \gamma_{ij}({\rho_H} - 
{{S_l}^l})\right)\right]
\eeqa
and constraint equations
\beqa
R + K^2  -K_{ij}~K^{ij} &=& 2  {\rho_{H}}, \\
 D_j{{K_i}^{j}} -  D_iK  &= & {J_i}.
\eeqa
Here $D_{i}$ is the covariant derivative operator with respect to
$\gamma_{ij}$, $R_{ij}$ is the three-dimensional Ricci tensor, 
$K$ is the trace of the extrinsic curvature, 
and $\gamma_{ij,t}$ denotes the partial derivative of
$\gamma_{ij}$ with respect to $t$. 
${\rho_{H}}$, ${J_i}$ and ${S_{i j}}$
 are defined by the right-hand-side of Eq.(\ref{einstein})  
(denoted as $T_{\mu\nu}$)  as 
\begin{eqnarray}
{\rho_{H}} & =&  T_{\mu\nu}n^{\mu}n^{\nu}, \\
{J_i}      & =&  -T_{\mu\nu}n^{\mu} {h_{i}}^{\nu}, \\
{S_{i j}}  & =&  T_{\mu\nu}{h_{i}}^{\mu} {h_{j}}^{\nu},
\end{eqnarray}
where 
\beq
h_{\mu\nu}=g_{\mu\nu}+n_{\mu}n_{\nu}
\eeq
is the projection operator on these hypersurfaces. We also need to
write the equation of motion of dilaton Eq.(\ref{scalar}) in the
first-order form. To do so, we introduce the following variable
likewise $K_{ij}$\cite{sst}
\beq
\Pi=-{\pounds}_n\phi=-{1\over
\alpha}\left(\phi_{,t}-\beta^i \phi_{,i}\right).
\eeq
Then the dilaton equation of motion can be written in the first-order 
form as
\beqa
\phi_{,t}-\beta^i\phi_{,i}&=&-\alpha \Pi,\\
\Pi_{,t}-\beta^i\Pi_{,i}&=& \alpha\left( K\Pi -\Pi^2
+(D\phi)^2-D^2\phi\right)-\gamma^{ij}\alpha_{,i}\phi_{,j}.
\eeqa

\subsection{Basic Equations in Spherical Symmetry}

Now we specialize the above general considerations to a 
spherically symmetric space-time. 
The most general line element of a 
spherically symmetric space-time is written in the form
\beq
ds^2=-(\alpha^2-A^2{\beta}^2)dt^2+2A^2\beta dtdr +
A^2dr^2+B^2r^2d\Omega^2,
\eeq
where $\beta $ is the radial component of the shift vector.
Because of the spherical symmetry, only two components of $K_{ij}$
are regarded as independent variables. We introduce $K_1\equiv {K^r}_r$. 
Because there does not appear ``bare'' $\phi$ in Eqs.(\ref{einstein}) and
(\ref{scalar}), it is convenient to introduce the radial derivative of it as 
another independent variable:
\beq
\Phi= {\phi_{,r}\over r}.
\eeq

Then the evolution equations in (3+1) form are written as
\beqa
A_{,t}-\beta A_{,r}&=&-\alpha A K_1+\beta_{,r}A,\label{evoa}\\
B_{,t}-\beta B_{,r}&=& -{1\over 2}\alpha B(K-K_1)+{\beta\over r} B,
\label{evob}\\
K_{,t}-\beta K_{,r}&=& -{1\over A^2}\left(4x \alpha_{,xx}+6\alpha_{,x}+
4x\alpha_{,x}\left(2{B_{,x}\over B}-{A_{,x}\over
A}\right)\right)\nonumber\\
&+& \alpha \left[  {1\over 2}K^2-KK_1+{3\over 2}K_1^2 + K\Pi -
\Pi^2+{x\Phi^2\over A^2}\right.\nonumber\\
&-& \left.{1\over A^2}\left( \left(4{xB_{,x}\over B}-2{xA_{,x}\over
A}+3\right)\Phi+2x\Phi_{,x}\right)
\right],\label{evok}\\
K_{1,t}-\beta K_{1,r}&=& -{1\over A^2}\left(4x \alpha_{,xx}+2\alpha_{,x}
-4x{A_{,x}\over A}\alpha_{,x}\right)\nonumber\\
&+&\alpha  \left[ {4\over A^2}\left( -2x{B_{,xx}\over B}-3{B_{,x}\over 
B}+{A_{,x}\over A}+2{A_{,x}B_{,x}\over AB}\right)\right.\nonumber\\
&+&\left. KK_1-K_1\Pi +{1\over A^2}\left(2x\Phi_{,x}+\Phi - 2x{A_{,x}\over
A}\Phi\right)
\right],\label{evok1}\\
\Pi_{,t}-\beta \Pi_{,r}&=& \alpha K\Pi -\alpha\left(\Pi^2 -{x\Phi^2\over
A^2}\right)\nonumber\\
&-&{\alpha \over A^2}\left(2{x\alpha_{,x}\over
\alpha}-2{xA_{,x}\over A}+4{xB_{,x}\over B}+3\right)\Phi
-2{x\alpha\over A^2}\Phi_{,x},\label{evopi}\\
\Phi_{,t}-\beta \Phi_{,r}&=& -2\alpha_{,x}\Pi -2\alpha \Pi_{,x}
+\left(\beta_{,r}+{\beta\over r}\right)\Phi,\label{evophi}
\eeqa
where we have set $x=r^2$ to include the regularity condition at the
origin. There are also constraint equations
\beqa
{2\over A^2}\left[  4{A_{,x}\over A}\right.&-&8{xB_{,xx}\over
B}-16{B_{,x}\over B}-4{xB_{,x}^2\over B^2}+8{xA_{,x}B_{,x}\over AB}
+{A^2\over xB^2}-
\left.{1\over x}\right] +{1\over 2}K^2 +KK_1 -{3\over 2}K_1^2\nonumber\\
&=& 2K\Pi -\Pi^2+x{\Phi^2\over A^2} - {2\over A^2}\left
( \left(4{xB_{,x}\over B}-2{xA_{,x}\over A}+
3\right)\Phi + 2x\Phi_{,x}\right)
\label{const:h}\\
K_{1,x}-K_{,x}&+&\left({B_{,x}\over B}+{1\over
2x}\right)\left(3K_1-K\right)
=-\Pi_{,x}+{1\over
2}K_1\Phi.\label{const:mom}
\eeqa

At this stage, it may be instructive to recover the homogeneous and
isotropic solutions. This can be done by taking $\alpha=1,\beta=0,
A=B$ and neglecting the spatial dependence in the variables. 
Then we arrive at the following equations:
\beqa
{A_{,t}\over A}&=&-{1\over 3} K\label{hubble}\\
K_{,t}&=& {1\over 3}K^2+K\Pi -\Pi^2\label{evo1}\\
\Pi_{,t}&=& K\Pi -\Pi^2\label{evo2}\\
{2\over 3}K^2&=&2K\Pi -\Pi^2,\label{frw}
\eeqa
with $\Pi=-\phi_{,t}$. By rewriting as $H\equiv -K/3$,
Eq.(\ref{hubble}) is 
nothing but the definition of the Hubble parameter. Eq.(\ref{frw}) is
the Friedmann equation. Eq.(\ref{evo1}) and Eq.(\ref{evo2}) are the
evolution equations. Either of them is redundant. Eq.(\ref{frw}) admits 
the following solution:
\beq
K=-3H={3\pm \sqrt{3}\over 2}\Pi.
\label{frw2}
\eeq
It is easily found that minus sign corresponds to the PBB branch  
solution, while positive sign corresponds to the post-big-bang branch
solution.

\section{numerical results}

\subsection{Initial Condition}

Initial data must satisfy the constraint equations Eq.(\ref{const:h})
and Eq.(\ref{const:mom}). 
Because of spherical symmetry, we can consider, without loss of
generality, the conformally flat three-metric:
\beq
d\l^2=\gamma_{ij}dx^idx^j= \psi(r)^4(dr^2+ r^2d\Omega^2).
\eeq
As a choice of the initial time slice(or choice of the extrinsic
curvature), we adopt the following extrinsic curvature  considering
the similarity with the homogeneous universe
\beq
K_{ij}={1\over 3}K \gamma_{ij}+\sigma_{ij}={1\over 2}(1-{1\over
\sqrt{3}})\Pi\gamma_{ij},
\label{slice}
\eeq
where $\sigma_{ij}$ is the traceless part of the extrinsic
curvature. That is, we take $\sigma_{ij}=0$ and $K=(3-\sqrt{3})\Pi/2$. 
The former condition is for simplicity while the latter condition is
chosen so that it corresponds to the Friedmann equation in the 
PBB phase in the homogeneous universe (see Eq.(\ref{frw2})). 

Then the constraint equations become
\beqa
4x\psi_{,xx}+6\psi_{,x}&=&x\psi_{,x}\Phi+{3\over 4}\psi\Phi+{x\over
2}\psi\Phi_{,x}-{x\over 8}\psi \Phi^2,\label{init:h}\\
\Pi_{,x}&=&{\sqrt{3}-1\over 4}\Phi \Pi.
\label{init:mom}
\eeqa
Note that because of the choice Eq.(\ref{slice}) the Hamiltonian 
constraint Eq.(\ref{init:h}) is now
independent of $K_{ij}$ and $\Pi$. Given $\Phi$, we solve
Eq.(\ref{init:h}) iteratively to get $\psi$ by the cyclic reduction 
method for a tridiagonal matrix.
However, the right-hand-side source
term does not have an appropriate $\psi$-dependence for numerical
treatment: as $\psi$ increases, the source term also increases, and
therefore the equation is numerically unstable. So, we ``redefine'' 
$\Phi$ by $\Phi_I \equiv \psi \Phi$, and rewrite the right-hand-side
of Eq.(\ref{init:h}) in terms of $\Phi_I$. That is,
\beq
4x\psi_{,xx}+6\psi_{,x}={x\psi_{,x}\over 2\psi}\Phi_I+
{3\over 4}\Phi_I+{x\over 2}\Phi_{I,x}-{x\over 8\psi} \Phi_I^2.
\eeq
Physically, fixing $\Phi_I$ corresponds (roughly) to fixing the proper 
kinetic energy density of the dilaton field. 
The boundary conditions are 
\beq
\psi_{,r}=0
\label{regular}
\eeq
at the origin for regularity and 
\beq
\left(r(\psi-1)\right)_{,r}=0
\label{robin}
\eeq
at the numerical outer boundary to guarantee the asymptotically flat
Friedmann property. Eq.(\ref{robin}) corresponds to setting the
asymptotic scale factor to be unity. 
We then solve Eq.(\ref{init:mom}) to get $\Pi$.

We consider the following Gaussian form for $\Phi_I$:
\beq
\Phi_I= -{\phi_0\over \Delta^2}\exp\left(-{x\over \Delta^2}
\right) .
\eeq
We examine whether changing the scale of inhomogeneity (i.e. $\Delta$) 
can affect the onset or the duration of the PBB inflation.
We use the unit where $H_i(r\rightarrow \infty) =-K_i(r\rightarrow
\infty)/3 =1$ initially, which determines the
time scale $t_0$ to the big bang singularity (in FRW universe) such that
$t_0=H_i^{-1}/\sqrt{3}$. 
The normalization factor $\phi_0$ is arbitrary and we set it to one.

\subsection{Numerical Details}

We describe the numerical details for solving the field equations by
the finite difference method. Time steps are labeled by the index $n$
and spatial grid points are labeled by $i$. 
We define the time derivative operator
\beq
(Z_{,t})^n_i\equiv {\Delta t_{n-1}\over \Delta t_n+\Delta t_{n-1}}
{Z^{n+1}_i- Z^n_i\over \Delta t_n} +{\Delta t_{n}\over \Delta t_n+
\Delta t_{n-1}}{Z^{n}_i- Z^{n-1}_i\over \Delta t_{n-1}} ,
\eeq
where $\Delta t_n\equiv t_{n+1}-t_n$ and a operator for a derivative
with respect to $x=r^2$
\beq
(Z_{,x})^n_i\equiv {x_i-x_{i-1}\over x_{i+1}-x_i}(Z_{i+1}^n-Z_i^n)+
{x_{i+1}-x_{i}\over x_{i}-x_{i-1}}(Z_{i}^n-Z_{i-1}^n).
\eeq
Then for a grid uniform in $r$, the evolution equations are 
second-order accurate in both space and time. 
We take a nonuniform grid because we need to perform numerical calculations 
with various initial conditions and therefore it is desirable to use 
small grid points. Typically we use 512 spatial grid points in this
work.  The time step is constrained by the Courant condition. 
For a nonuniform grid, we use
\beq 
\Delta t = \varepsilon_1 \min_i\left( {r_{i+1}-r_i\over \beta_i \pm
2\alpha_i/A_i}\right),
\eeq
and we take $\varepsilon_1 =0.2$ for accuracy. 
The Courant
condition is sufficient for stability, however,  not for accuracy. 
Therefore we impose the additional requirement that $A,B,\Phi$ do
not change too much (near origin) in one time step:
\beq
dt=\varepsilon_2\min\left({A\over A_{,t}},{B\over B_{,t}},
{\Phi\over \Phi_{,t}}\right),
\eeq
and choice of the value of $\varepsilon_2$ depends on the initial
condition. For example, we choose $\varepsilon_2 =10^{-3}$ for
$\Delta/H_i^{-1}=1.0$.

Regarding the boundary conditions, Eq.(\ref{const:mom}) requires a
single boundary condition. We set $K_1=K/3$ at the origin for
regularity. Evolution
equations Eq.(\ref{evok})-Eq.(\ref{evophi}) require an outer boundary
condition as well as the regularity condition at the origin of kind 
Eq.(\ref{regular}). To determine the value of a variable at $i_{max}$,  
we extrapolate linearly in $x$ using the values of the 
variable at $i_{max}-1$ and $i_{max}-2$. This is because it is not
obvious that the conventional outgoing-wave boundary condition
 should be imposed in the present context. 
Since we locate the numerical
outer boundary at the point much further than the scale of 
inhomogeneity $( > 10 \times \Delta)$ where the universe is 
approximately FRW, such a choice is sufficient in practice. 

\subsection{Numerical Test}

As a test of the numerical code, we have used homogeneous initial
conditions and compared the numerical solutions with the exact
homogeneous solutions. The results are shown in Fig.1. There the exact 
solutions are shown as solid curves, while the numerical solutions at
the center ($r=0$) are plotted as crossed points. 
We find excellent agreement. 
In fact, the maximum relative error between the numerical solution and
the exact solution was less than $10^{-4}$ and the maximum relative
error in the Hamiltonian constraint was less than $10^{-8}$ because
the variables appear only quadratically in the constraint. 

Barrow and Kunze derived an exact solution of the collapse of
a homogeneous spherical region of the stiff ($p=\rho$) fluid \cite{bk}. 
The comparison with their solution is difficult in the present work 
because the spatial distribution of the dilaton is step-function-like
and hence $\Phi$ becomes singular near the surface
(delta-function-like). Furthermore, Eq.(\ref{evopi}) contains the
spatial derivative of $\Phi$, which develops a further ill-behaved
singularity.

\subsection{Results of Numerical Calculations}

We choose the geodesic slicing condition, $\alpha=1$ with zero shift, 
$\beta=0$.  We use Eqs.(\ref{evoa}-\ref{evok}) and 
(\ref{evopi}-\ref{evophi}) to solve $A,B,K,\Pi,$ and $\Phi$ and use 
the momentum constraint Eq.(\ref{const:mom}) instead of
Eq.(\ref{evok1}) to solve $K_1$. The Hamiltonian constraint 
equation Eq.(\ref{const:h}) is used as a check of the numerical
accuracy. 
As a geometrically invariant diagnostics for the inhomogeneity, we  
calculate the Bel-Robinson tensor\cite{bel}. The Bel-Robinson tensor 
is defined in terms of the Weyl tensor as
\beq
 T_{\mu\nu\rho\sigma}\equiv
C_{\mu\alpha\rho\beta}{{{C_{\nu}}^{\alpha}}_{\sigma}}^{\beta}
+~ ^*C_{\mu\alpha\rho\beta}{{{ ^*C_{\nu}}^{\alpha}}_{\sigma}}^{\beta},
\eeq
where $C_{\mu\nu\rho\sigma}$ is the Weyl tensor and 
 $^*C_{\mu\nu\rho\sigma}\equiv {1\over
2}\epsilon_{\mu\nu\alpha\beta}{C^{\alpha\beta}}_{\rho\sigma}$ is its
dual. The ``superenergy'' density on the three-dimensional spacelike
hypersurface with $n^{\nu}$ being the unit normal vector is defined by
\beq
W \equiv T_{\mu\nu\rho\sigma}n^{\mu}n^{\nu}n^{\rho}n^{\sigma}
=E_{\mu\nu}E^{\mu\nu}+B_{\mu\nu}B^{\mu\nu},
\eeq
where 
\beq
E_{\mu\nu}\equiv C_{\mu\alpha\nu\beta}n^{\alpha}n^{\beta},~
B_{\mu\nu}\equiv ~^*C_{\mu\alpha\nu\beta}n^{\alpha}n^{\beta}.
\eeq
They can be written  in the present case as
\beqa
&&E_{ij} = R_{ij}+KK_{ij}-{K_i}^{l}K_{lj}
-{1\over 2}\left(S_{ij}-{1\over 3}\gamma_{ij}{S^l}_l\right)-
{2\over 3}\gamma_{ij}\rho_H,\\
&&B_{ij} =\epsilon_{lm(i}D^l{K^m}_{j)}.
\eeqa
We have numerically solved the field
equations with various choice of $\Delta$. Numerical results are
shown in Figs.2,3,4 for $\Delta/H_i^{-1}=1,5,0.2$. 

For $\Delta/H_i^{-1}=1$(Fig.2), which can be regarded as a fiducial case,  
the evolution was terminated at $t\simeq
0.49$ beyond which numerical calculations became inaccurate near the  
origin. The inhomogeneities in $\Phi$ and $\Pi$ always grows during
the evolution. The superenergy of the Bel-Robinson tensor 
shows that after the decay of the initial irregularities, 
new irregularities emerge and grow. This phenomena can be understood
as follows. Initially inhomogeneous region expands being dragged by the
cosmic expansion. However, its self gravitational energy dominates
over the background energy density soon, and then that lump decouples
from the cosmic expansion and begin to collapse. 
We note that although irregularities are
increasing, the universe nonetheless expands rapidly since the
expansion rate is increasing (See Fig.2(a)(b)) 
The situation is to be contrasted with the usual potential 
energy dominated inflation (for example, see Fig.14 in \cite{gp} and
Fig.3 in \cite{shinkai}).

For $\Delta/H_i^{-1}=5$(Fig.3), the evolution was terminated at $t\simeq
0.46$. Further evolution made the numerical calculations inaccurate
near the origin. We find that the behavior similar to
$\Delta/H_i^{-1}=1$ case appears again. 
For $\Delta/H_i^{-1}=0.2$(Fig.4), the evolution was terminated at $t\simeq
0.12$. Even if we perform numerical evolution further, the evolution
was forced to stop soon. Presumably the singularity is reached at the 
origin first. Note that $K$ and $K_1$ are growing positively near
origin. That means the expansion is delayed and eventually turns into 
the contraction to the singularity. Consequently, the metric (or scale 
factor) does not grow much near the origin. 
We do not see the growth of inhomogeneity in the Bel-Robinson tensor 
probably because there is no time for them to overcome the cosmic
expansion. If we could perform the numerical evolution any further
more accurate, we could see the same structure as in Fig.2 and Fig.3. 

We also calculate the e-folding number $N$ of the expansion at the origin
\beq
N=\ln \left\{\left(A(r=0)B(r=0)^2\right)^{1/3}/\psi(r=0)^2\right\},
\eeq
and the results are shown in Fig.5. For comparison we also
plotted the e-folding number for the FRW universe. We find that 
for $\Delta \siml 0.3 \times H^{-1}_i$ the universe does not inflate
sufficiently and/or recollapses. 
Note that the e-folding number for $\Delta \simg 0.4H_i^{-1}$, where the
onset of PBB-inflation is not prevented (in agreement with the rough
estimate in \cite{inhomo3}),  is larger than
that of the homogeneous solution. This is because initially
the inhomogeneous dilaton is superposed on the background homogeneous 
solution and that inhomogeneous energy contributes in turn to the
superinflationary expansion. In \cite{inhomo3}, a criterion for PBB
inflation is proposed that is translated roughly as, $\Delta \simg
H_i^{-1}$, which is consistent with our numerical results. 

Our results may be understood intuitively in the Einstein frame. In the
Einstein frame, the PBB phase corresponds to a recollapsing
universe with a massless scalar field rather than an expanding universe. 
So, inhomogeneities should grow. 

To conclude, we find that a large inhomogeneity, $\Delta \siml
0.3H^{-1}_i$, reduces the amount of the PBB inflation and sometimes
even does not allow the development of the PBB inflation as in the
case of the usual inflation\cite{gp2}. However, even if the 
universe enters the PBB inflation stage, the initial inhomogeneity
grows and is not smoothed out globally at all. 

\subsection{Fine-Tuning Problem}

We consider the meaning of the condition for the onset of the PBB
inflation($\Delta \simg 0.4H^{-1}_i$) from the viewpoint of the
resolution of the horizon problem\cite{finetune1}. 

Let the time of the beginning of the PBB inflation be $t_i$ and the
final time be $t_f$ and the time of the final singularity be $t_s$. 
$t_f$ is supposed to be the time when the stringy nonperturbative
effects become dominant and the low energy effective action
Eq.(\ref{action}) is no longer valid. The decrease of the
comoving Hubble length, $(Ha)^{-1}$, measures the amount of inflation.
Using the PBB solution in FRW background \cite{pbb,finetune1}, we have
\beq
Z\equiv {H_fa_f\over H_ia_i}=
\left({t_s-t_i\over t_s-t_f}\right)^{1+1/\sqrt{3}} 
\simeq \left({t_s-t_i\over l_s}\right)^{1+1/\sqrt{3}}, 
\label{inequality}
\eeq
where $a$ is the scale factor and $H_f^{-1} \simeq t_s-t_f \simeq l_s$ 
is assumed. $Z>e^{60}$ is required to solve the horizon problem
\cite{kt}. On the other hand, we find that the lump of the initial
size $\Delta \simg 0.4H^{-1}_i$ inflates. Using Eq.(\ref{inequality})
this inequality can be rewritten as
\beq
 \Delta \simg 0.4H^{-1}_i \simeq 0.4(t_s-t_i)> e^{37}l_s,
\eeq
which shows that initially extremely large inhomogeneous region (in the 
unit of the string scale) is required in order to get the PBB
inflation and to solve the horizon problem. This large number may be 
regarded as  fine-tuning of the initial condition because $l_s$ is a 
natural length scale in string theory.

\section{summary}

We have studied numerically the effect of the initial inhomogeneities
on the onset and duration of the PBB inflation. 
We found that a large  initial inhomogeneity does suppress the onset 
of the PBB inflation as in the case of the usual inflation\cite{gp2}. 
Further, even if the PBB inflation is realized, the initial 
inhomogeneity grows contrary to the usual inflation. 
The initial scale of inhomogeneity is required to be extremely large 
(in the unit of the string scale) to realize the PBB inflation and to
solve the horizon problem. 

The fine-tuning problem of the PBB inflation was recently addressed in 
\cite{finetune1,finetune2} for homogeneous cosmologies 
in the context of the horizon problem and the flatness problem: 
the pre-big-bang scenario is quite sensitive to spatial curvature and
anisotropy. Combining these results with our results, it may be said
that if the present universe would be resulted from the PBB inflation,
the universe would have to be initially extremely homogeneous and 
isotropic and flat. 

Of course, our study does not exhaust all the possibility; (i) we did
not include other massless degrees of freedom (for example, the
axion), (ii) the asymptotic condition was limited to the flat FRW
universe, (iii) the space-time dimensionality was fixed to four. 
It would be interesting to relax the above situations.

Hawking and Moss once said in \cite{gibbons} 
that the asymptotic approach
to the de Sitter state (in the usual inflation) is ``very similar to the 
way that a gravitational collapse rapidly approaches a stationary 
black-hole state (outside of the black-hole horizon) which depends 
only on the mass and angular 
momentum but which is otherwise independent of the nature of the
collapsing body''. By the same black-hole analogy, the PBB 
inflation may be similar to the ``interior'' of a black-hole which is
dependent on the nature of the collapsing body (initial conditions).

\acknowledgments
The author would like to thank Professor Kei-ichi Maeda for useful
discussions. He is also grateful to Professor Gabriele Veneziano for
useful correspondence. 
A part of this work was reported at the annual meeting of 
the physical society of Japan, March, 1998. 
The author is supported by a JSPS Research Fellowship for 
Young Scientists under grant No.3596.

\newpage
\vskip 0.05in
\centerline{FIGURE CAPTION}
\vskip 0.05in

\newcounter{fignum}
\begin{list}{Fig.\arabic{fignum}.}{\usecounter{fignum}}

\item
Solutions with homogeneous initial conditions. Solid curves 
correspond to the exact homogeneous solutions, while crossed points 
correspond to the numerical solutions. We find excellent agreement.

\item
Evolutions from inhomogeneous initial conditions up to $t\simeq 0.49\times 
H^{-1}_i$ for  $\Delta=1.0\times H^{-1}_i$. 
All variables are normalized by $H_i$. Thick curves are initial
values. Time evolution is in the 
direction from top to bottom for $K,K_1,\Phi,\Pi$ and the relative
error,  while bottom to top for $A$ and $B$. For the superenergy of
the Bel-Robinson tensor, first the initial lump decays, and then new lump
appears near the origin.

\item
Evolutions from inhomogeneous initial
conditions up to $t\simeq 0.46\times H^{-1}_i$ for 
$\Delta=5.0\times H^{-1}_i$. 
Thick curves are initial values. 

\item
Evolutions from inhomogeneous initial
conditions up to $t\simeq 0.12\times H^{-1}_i$ for 
$\Delta=0.20\times H^{-1}_i$. 
Thick curves are initial values. 

\item
E-folding number at the center as a function of
time. $\Delta/H^{-1}_i=5.0,1.0,0.5,0.4,0.3,0.2,0.1$ from top to bottom. 
Dashed line corresponds to the homogeneous solution. We find that for
$\Delta \siml 0.3 \times H^{-1}_i$ the universe does not inflate
sufficiently.

\end{list}

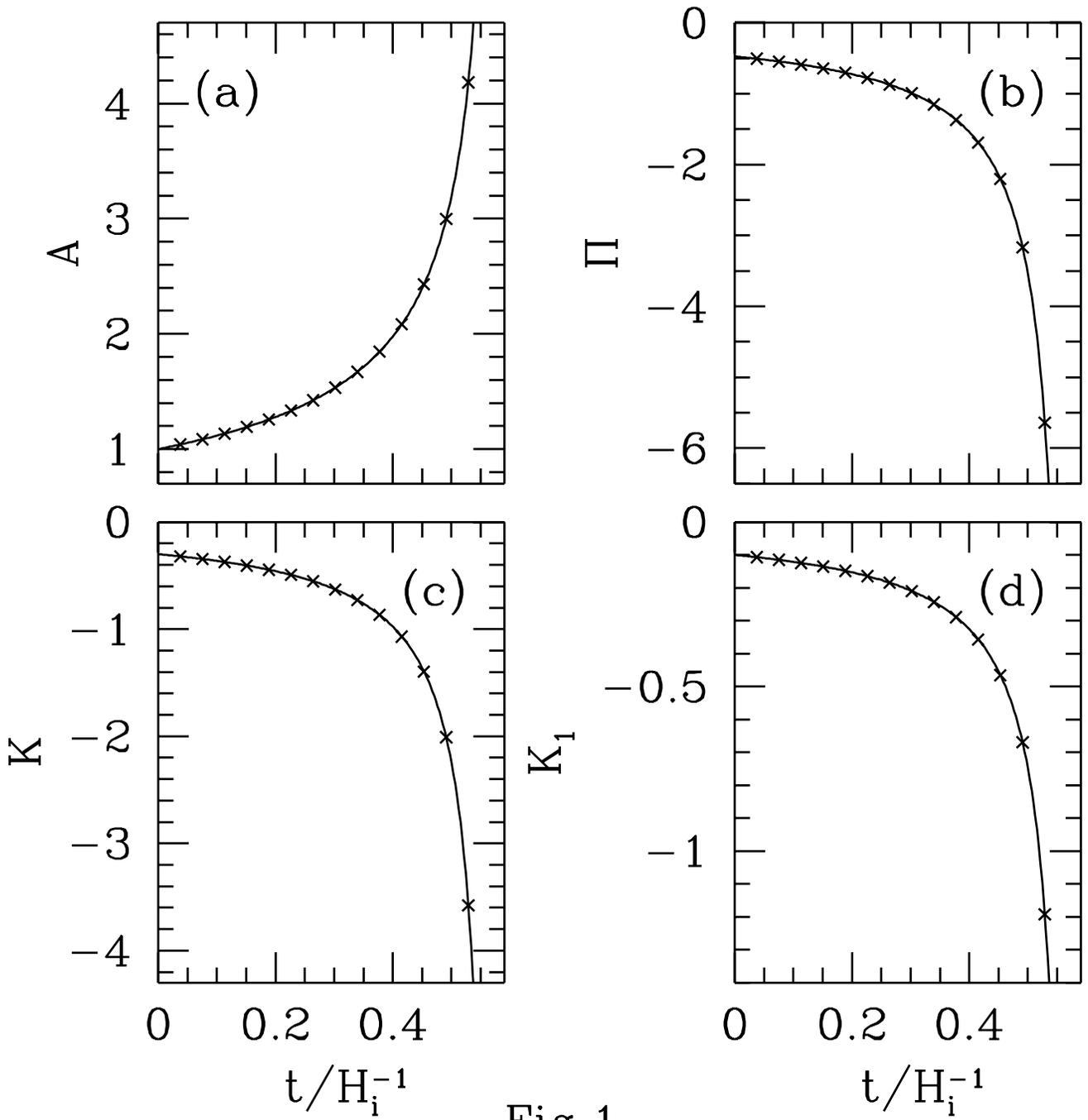
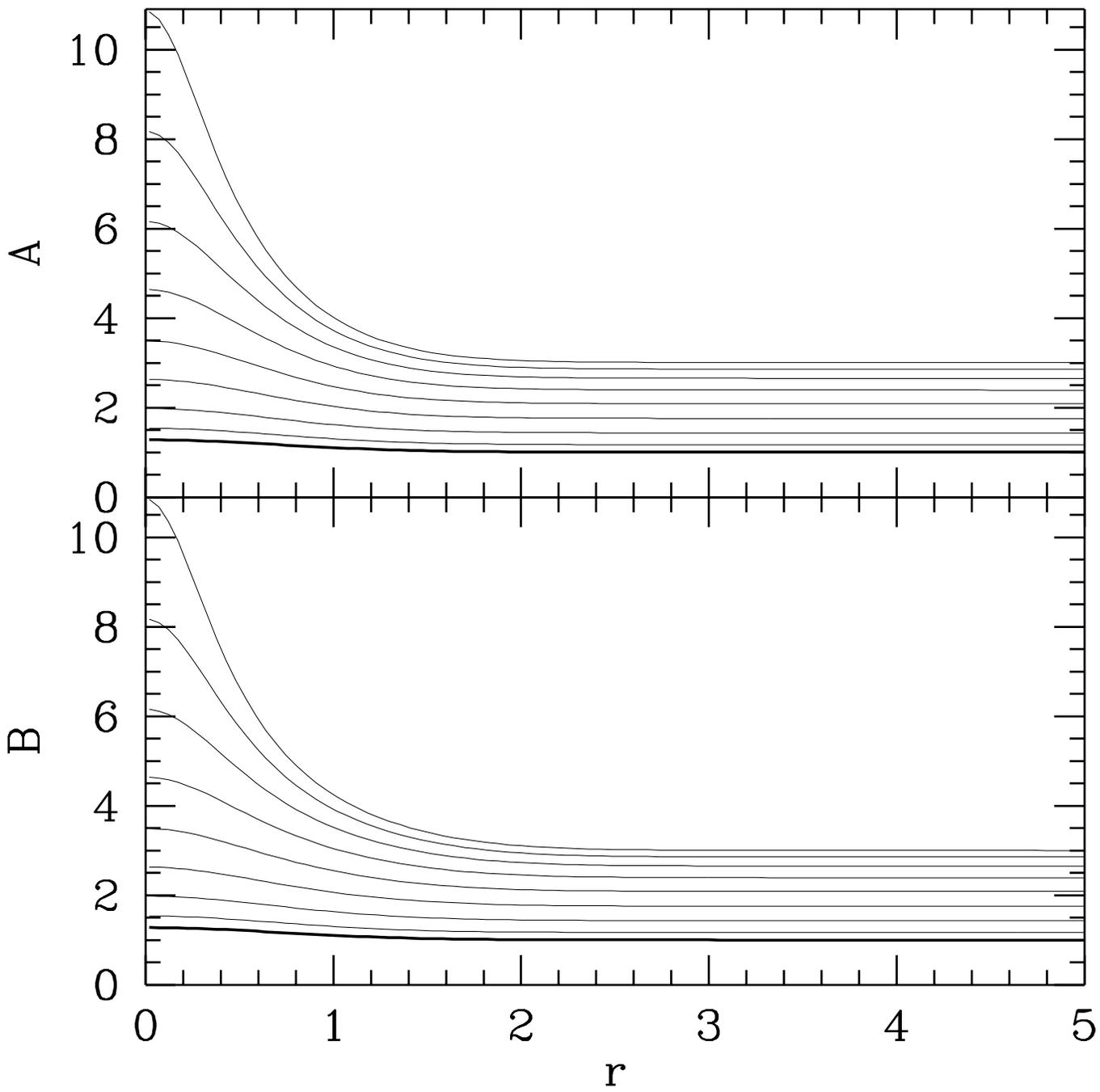
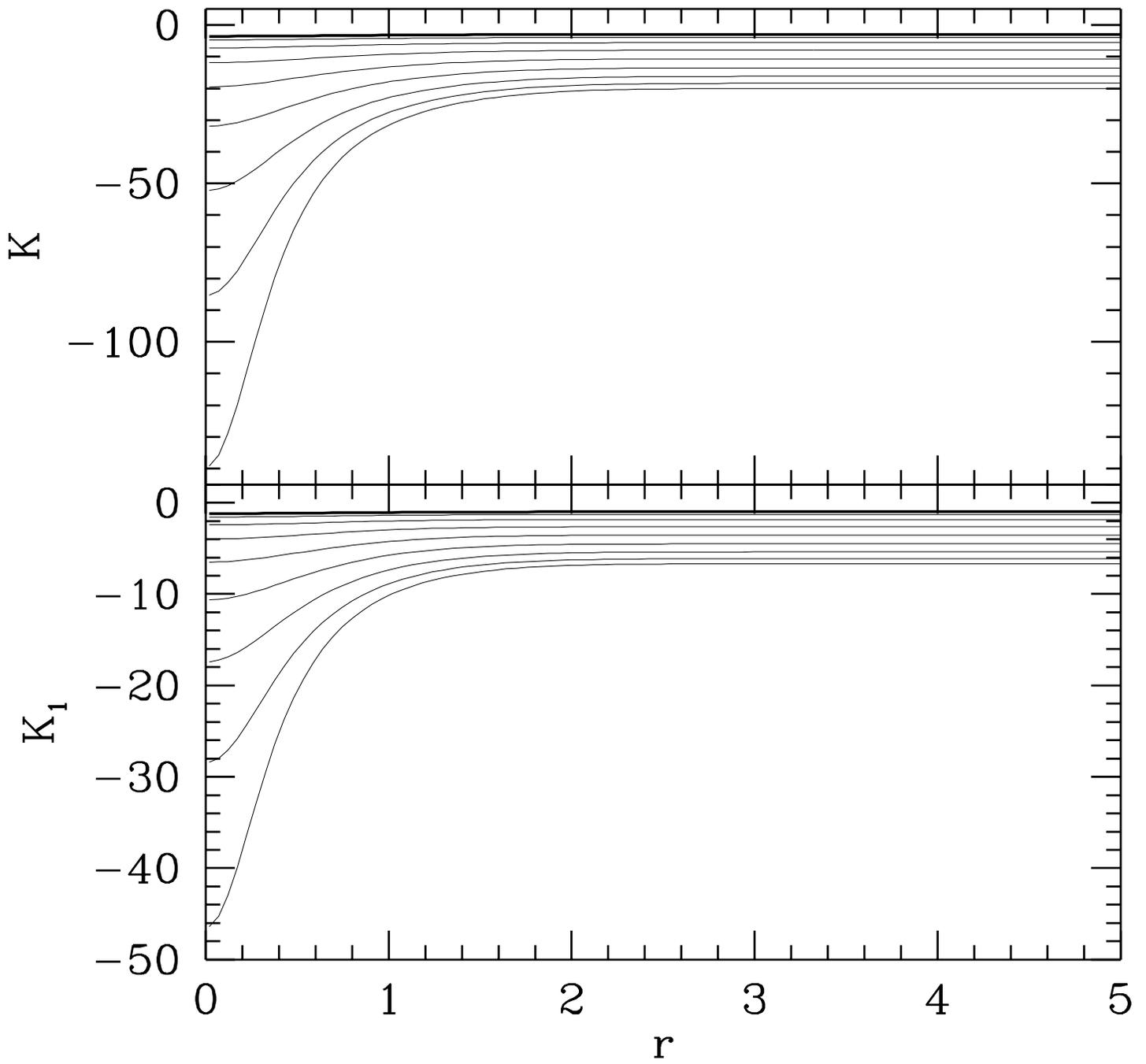
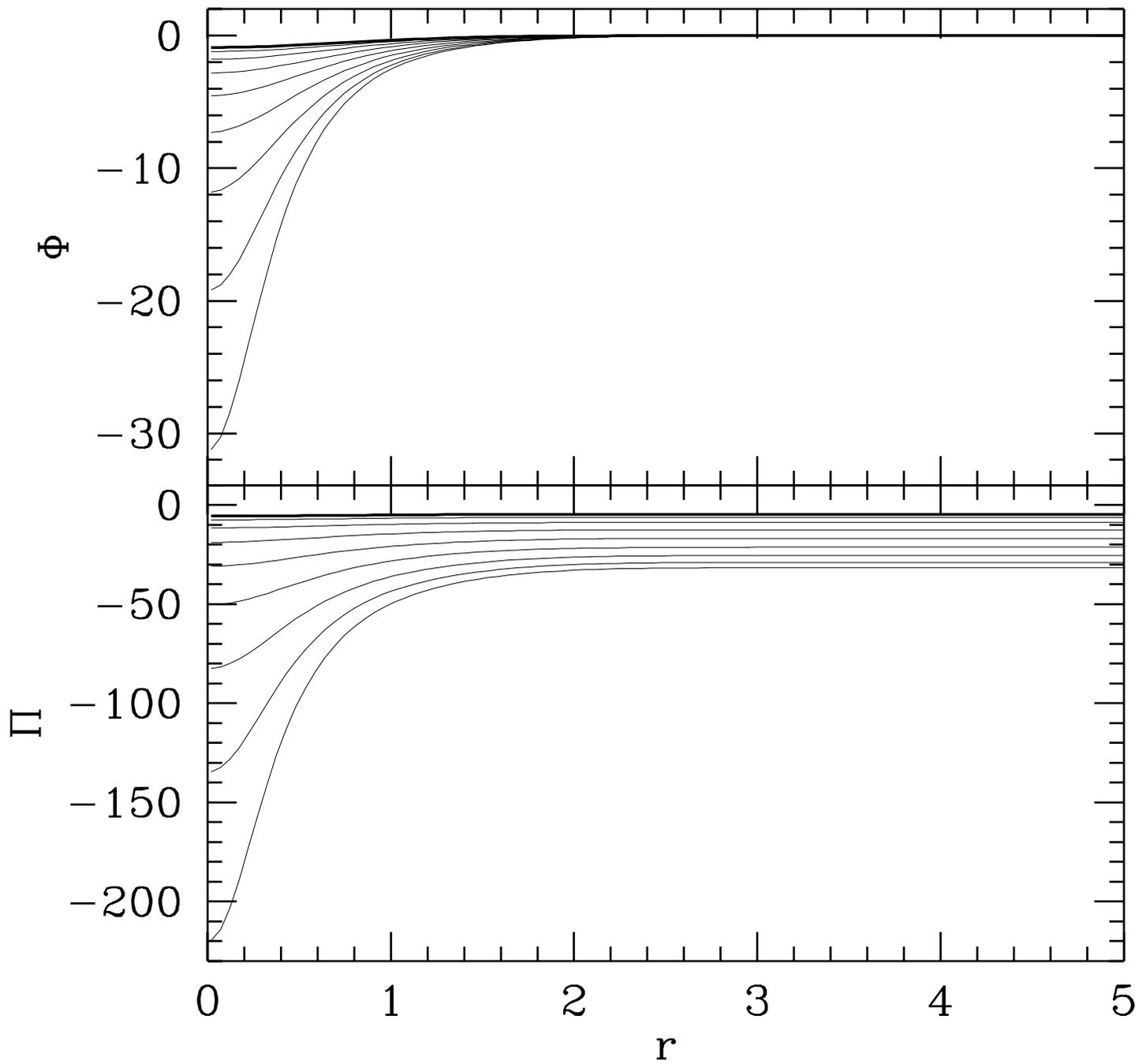
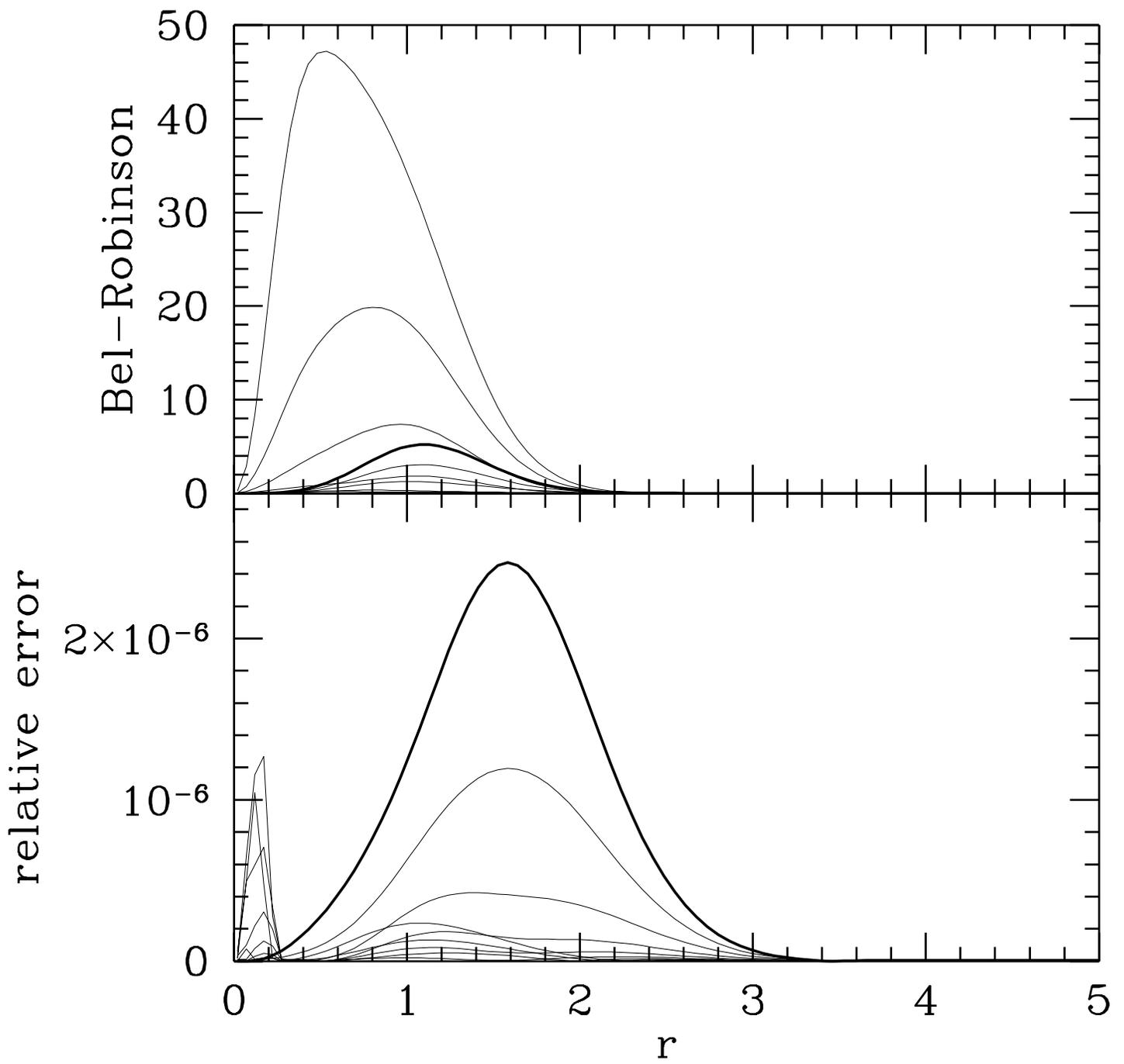
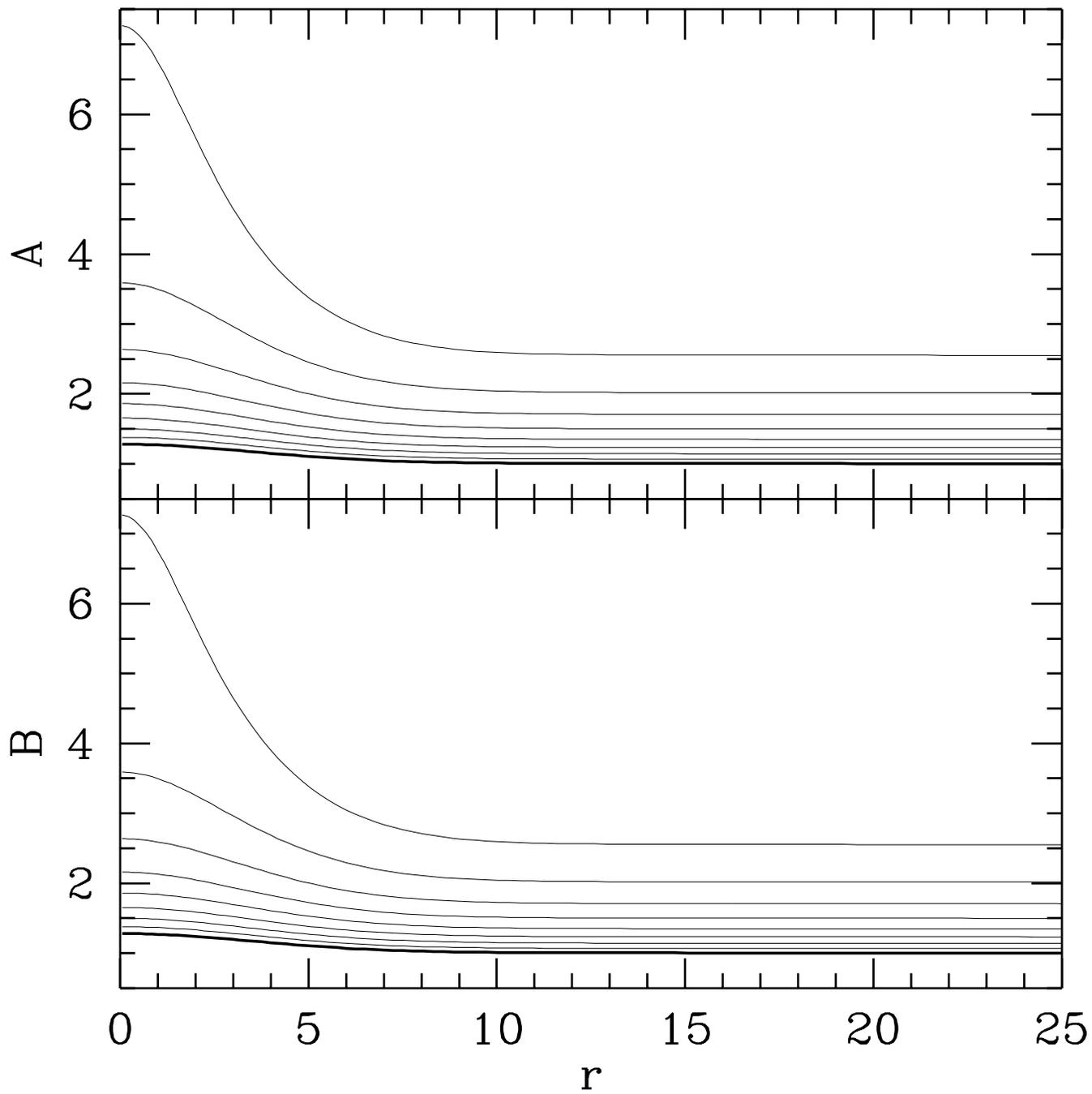
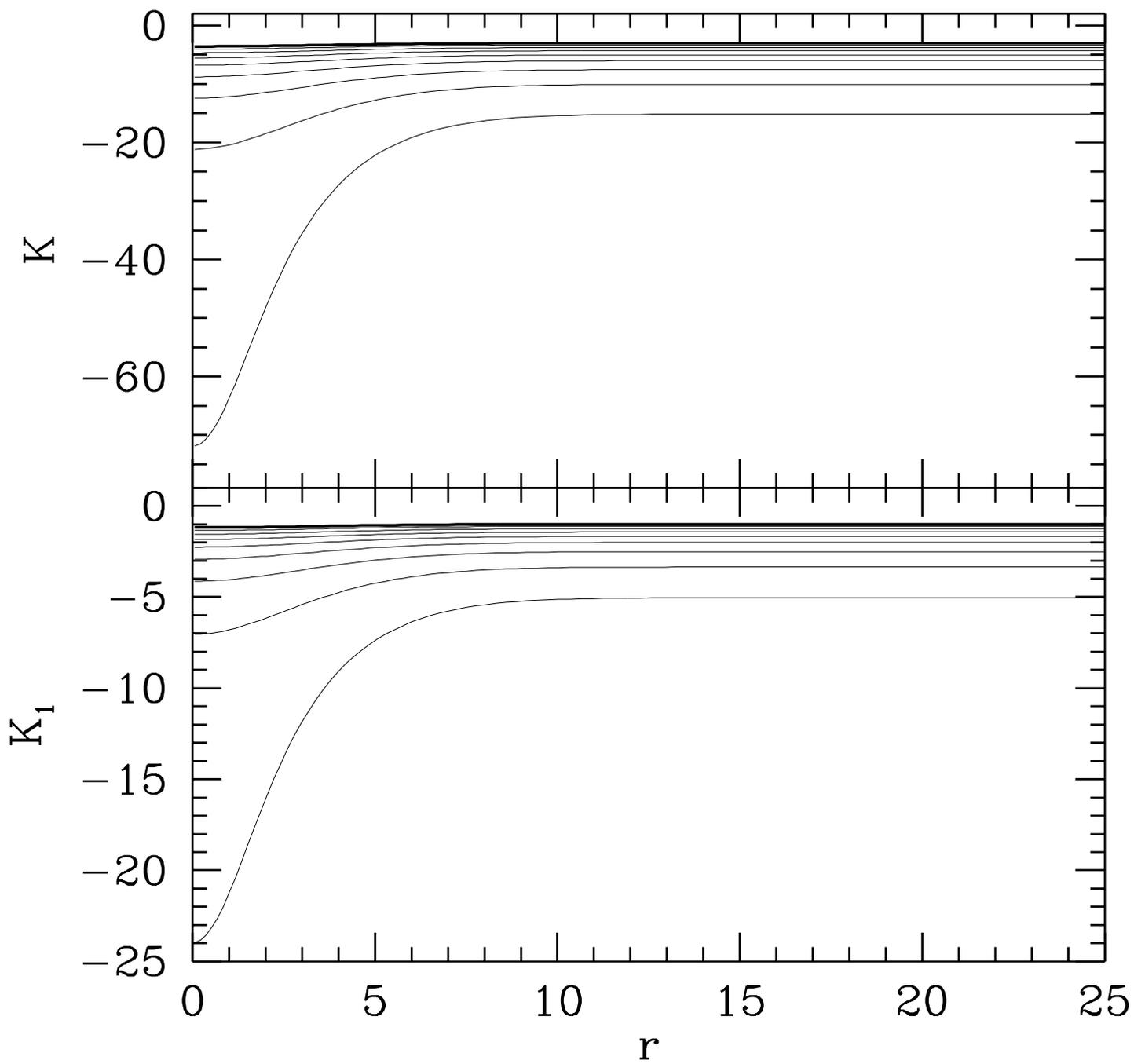
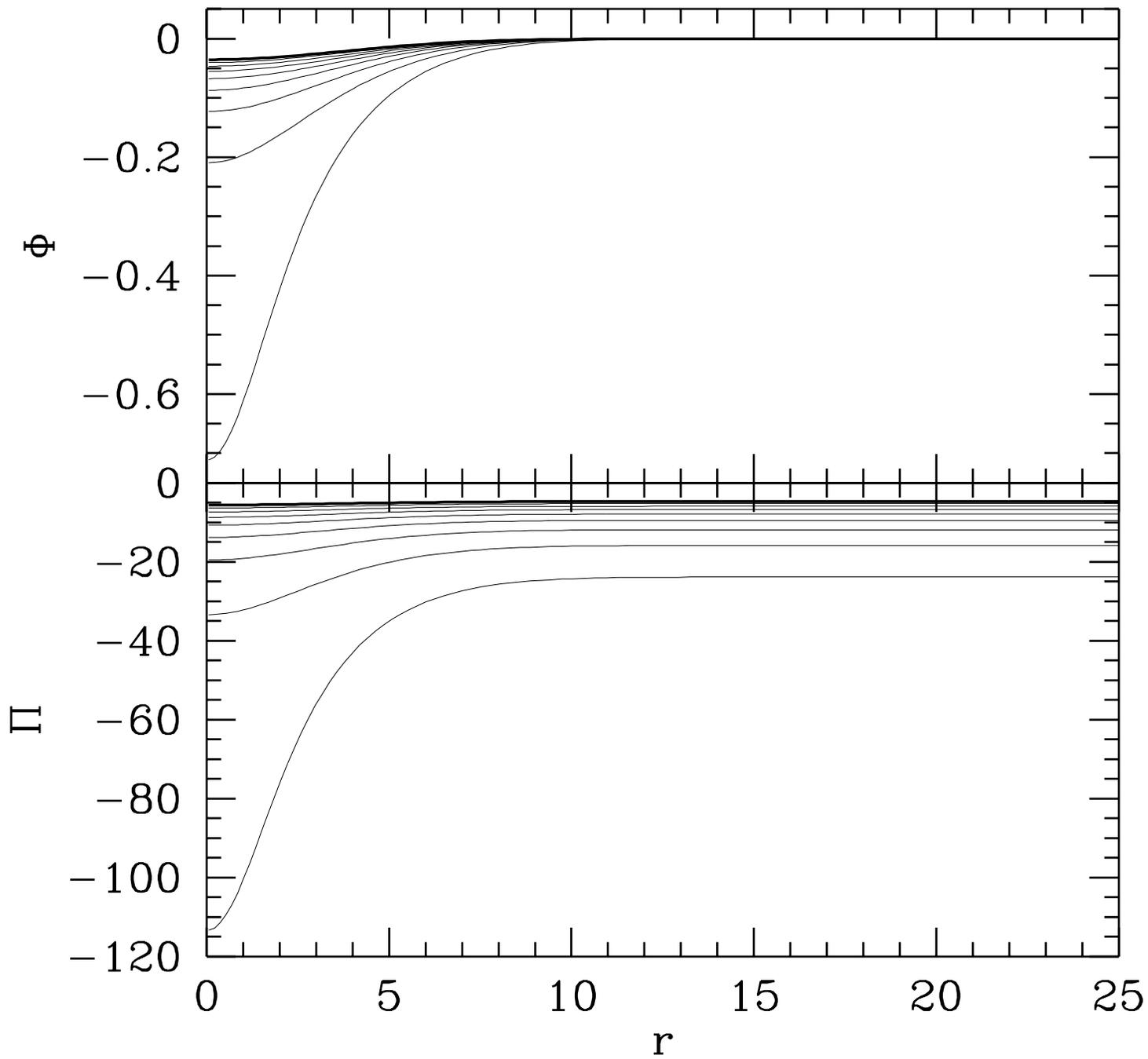
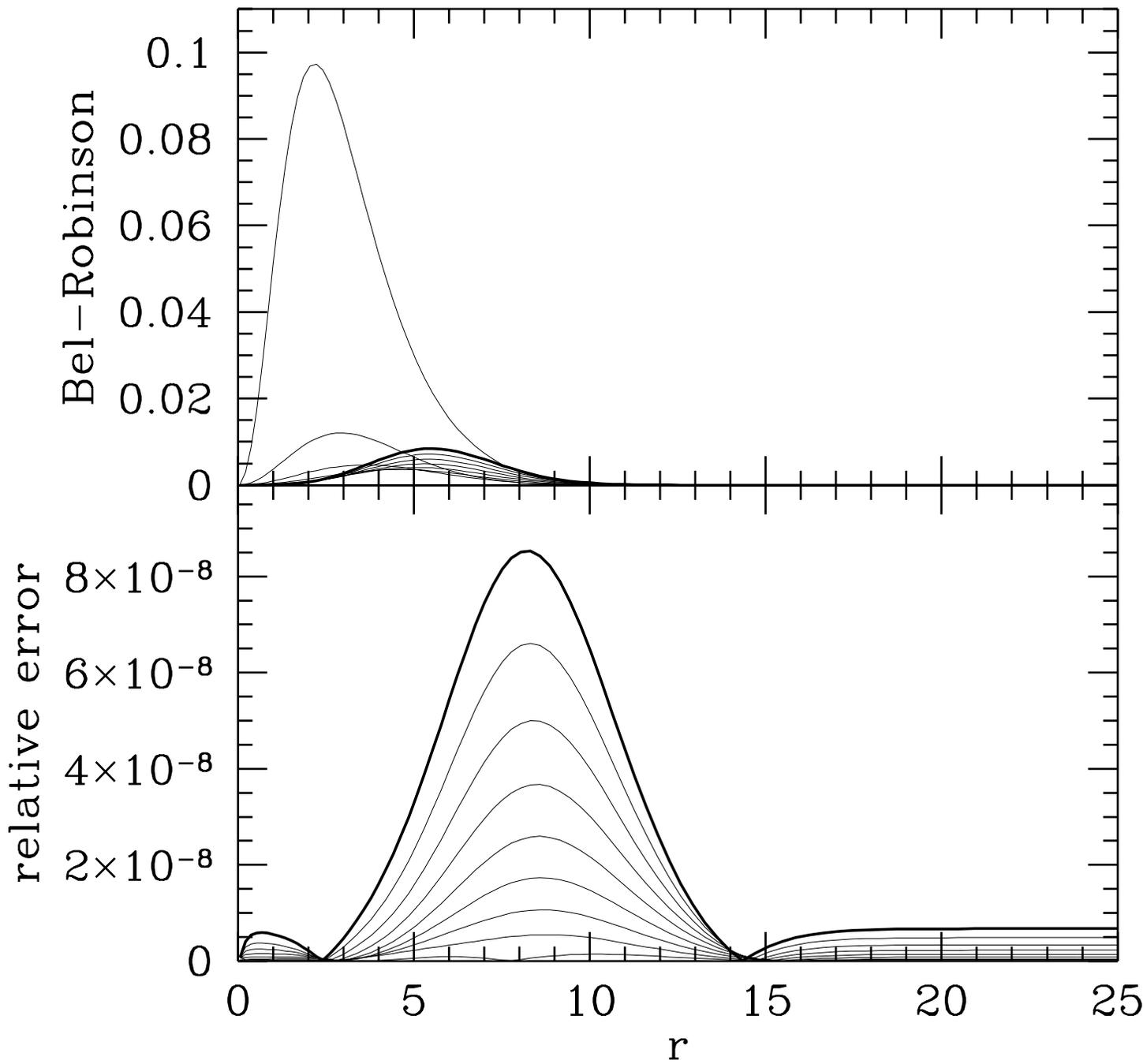
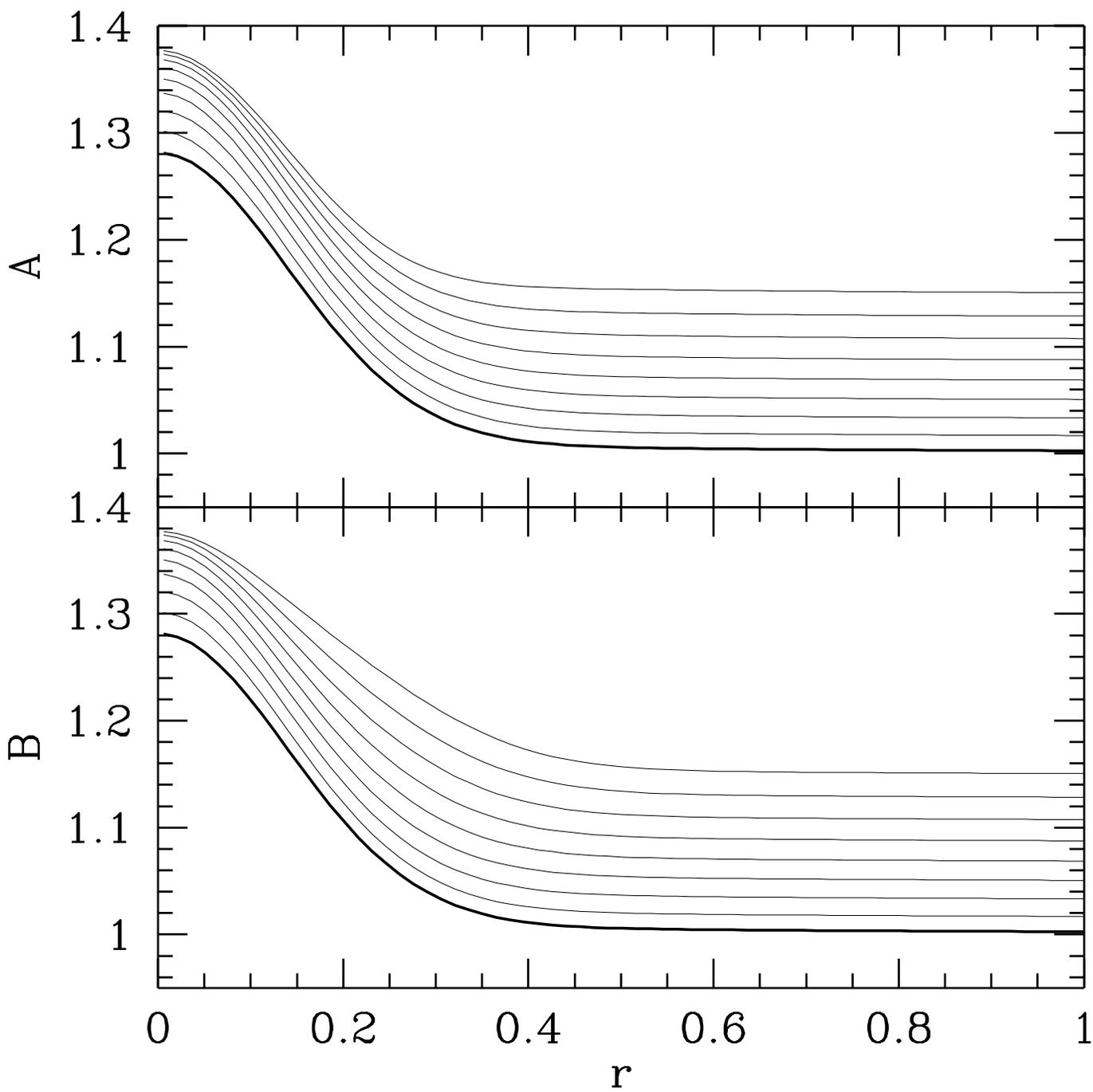
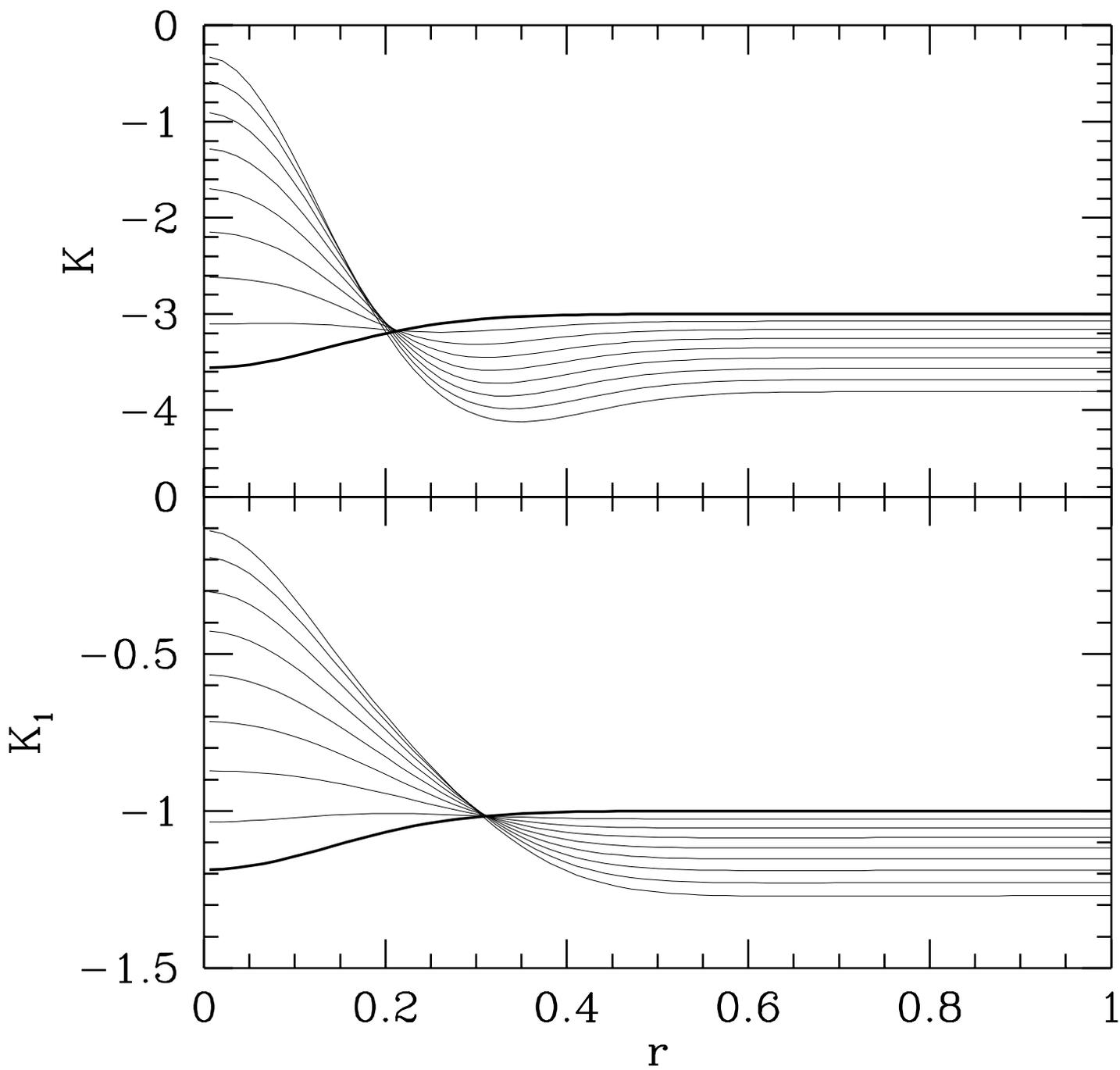
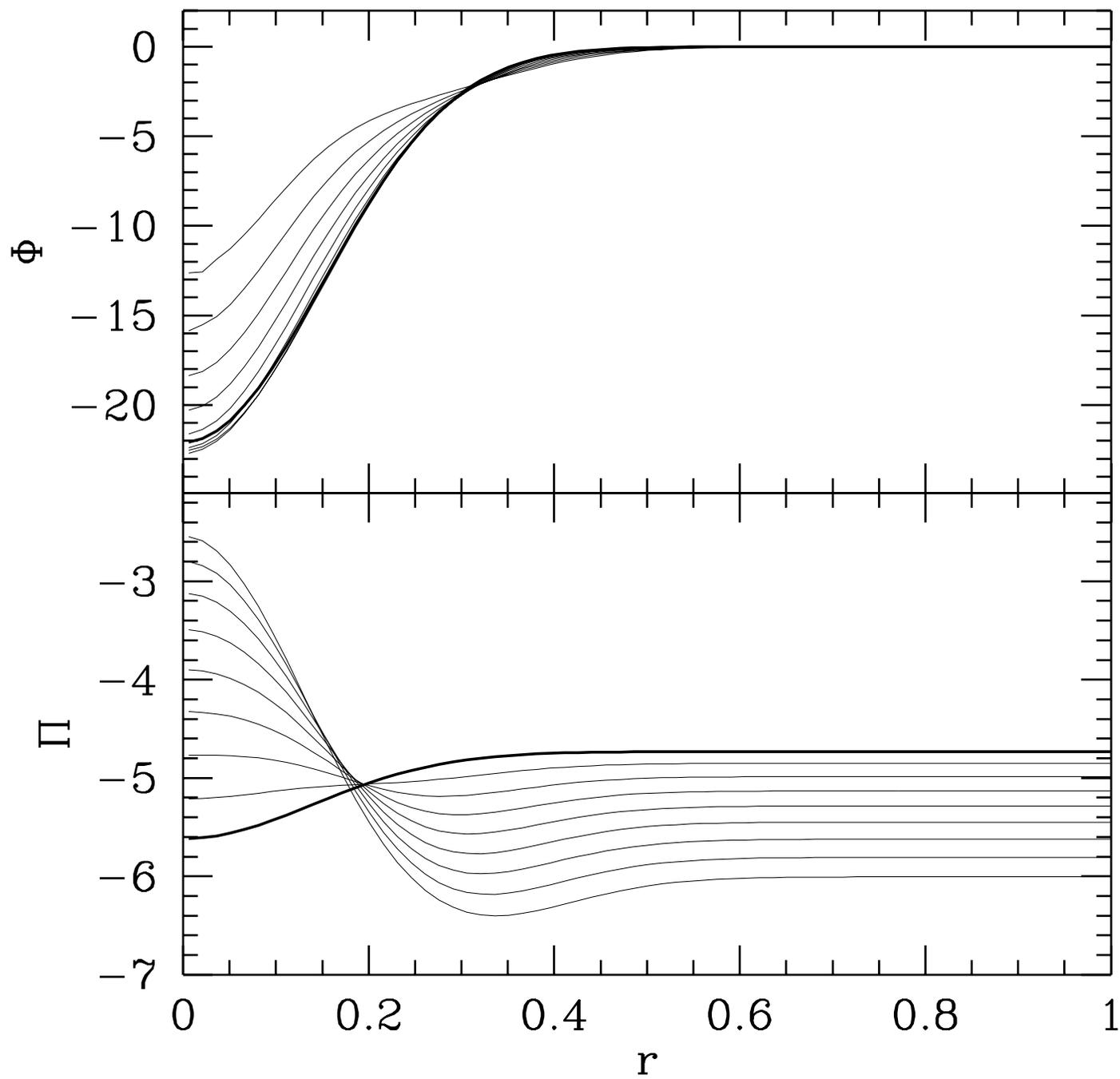
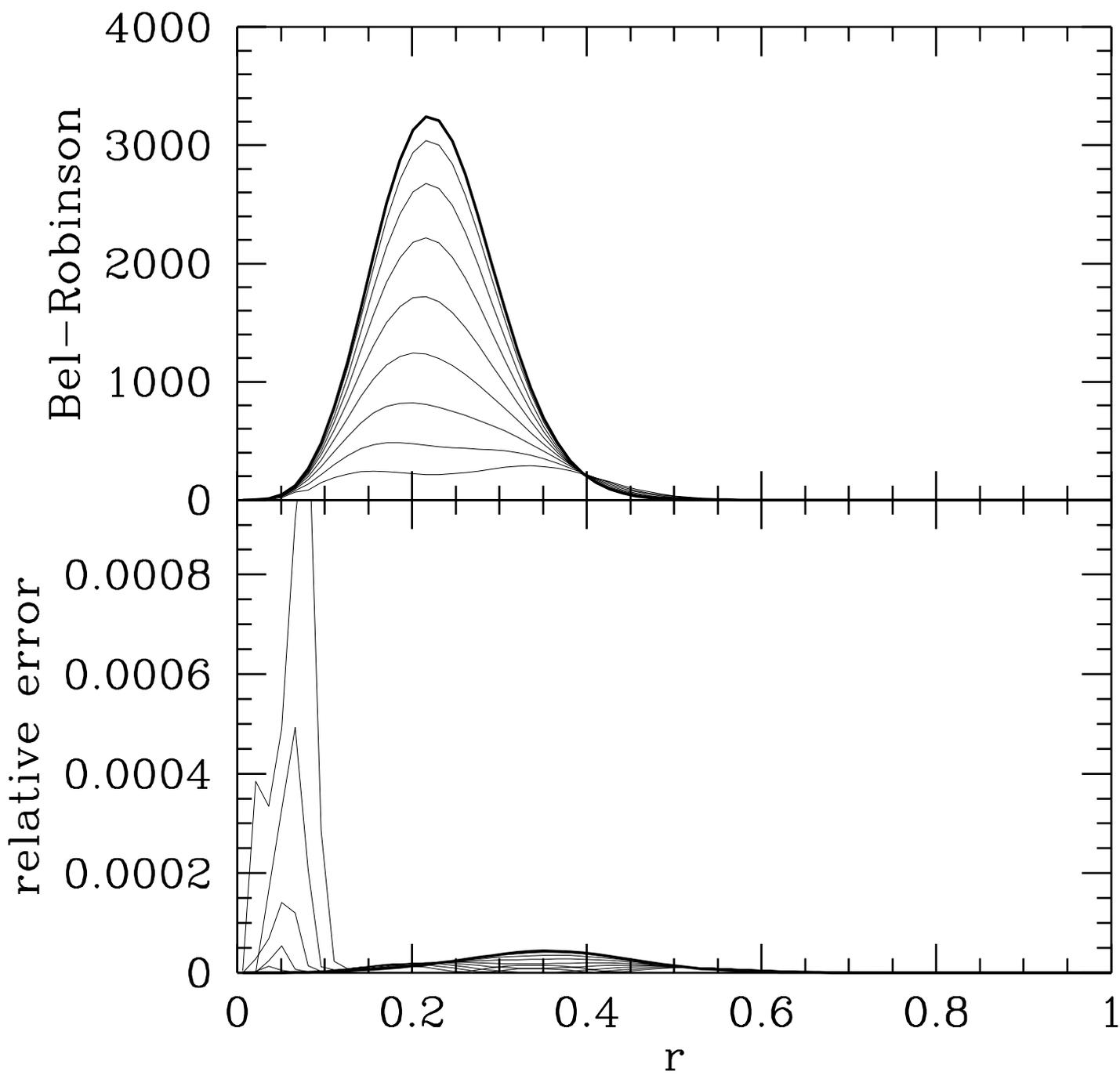
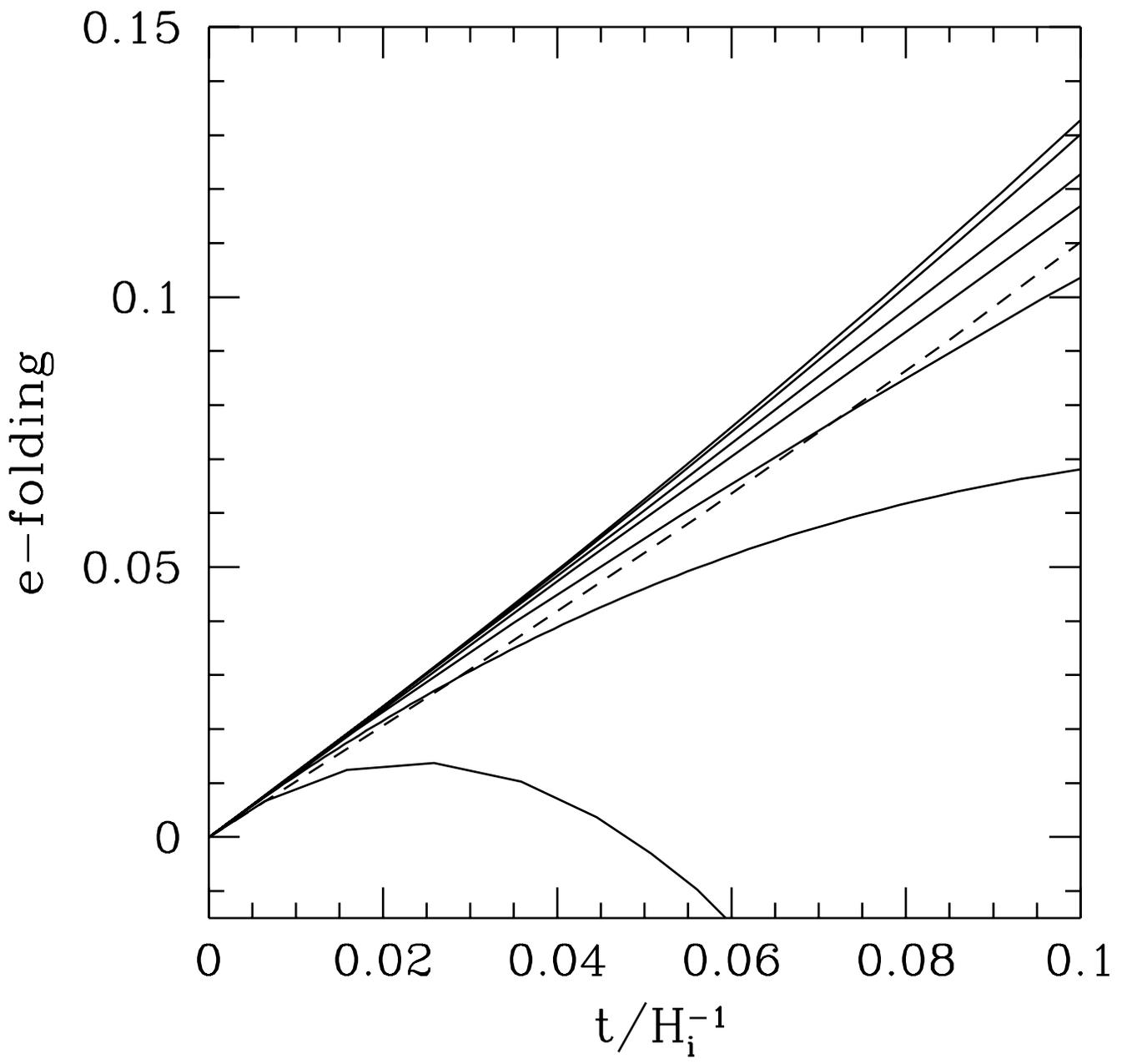
\end{document}